\documentclass[10pt,prd,aps,amssymb,amsmath]{revtex4}

\usepackage{natbib}
\usepackage{hyperref}
\usepackage{graphicx} 

\begin{document}


\title{ 
Yields: The Galápagos Syndrome Of Cryptofinance
}

\author{ Bernhard K. Meister$^{1}$ \& Henry C. W. Price$^{2,3}$}
\affiliation{ 
$^{1}$MK2 Finance, Vienna, Austria \\ 
$^{2}$Department of Physics, Imperial College London, London 
SW7 2AZ, UK \\ 
$^{3}$Centre for Complexity Science, Imperial College London, 
London SW7 2AZ, UK}
\email{bernhard.k.meister@gmail.com, henry.price10@imperial.ac.uk }


\begin{abstract}
\noindent
In this chapter structures that generate yield in cryptofinance will be analyzed and related to leverage. While the majority of crypto-assets do not have intrinsic yields in and of themselves, similar to cash holdings of fiat currency, revolutionary innovation  based on smart contracts,  which enable \emph{decentralised} finance, does generate return. Examples include lending or providing liquidity to an automated market maker on a decentralised exchange, as well as performing block formation in a proof of stake blockchain. 
On centralised exchanges, perpetual and finite duration futures can trade at a premium or discount to the spot market for extended periods with one side of the transaction earning a yield.  Disparities in yield exist between products and venues as a result of market segmentation and risk profile differences. 
Cryptofinance was initially  shunned by legacy finance and  developed independently. This led to curious and imaginative adaptions, reminiscent of Darwin's finches, including stable coins for dollar transfers, perpetuals for leverage, and a new class of exchanges for trading and investment.
 \end{abstract}
 
\maketitle


\section{Introduction:   
Decentralised Finance needs
Yield }
\noindent
The combined market capitalisation of cryptocurrencies has grown exponentially  
and stands as of June 2021 at around \$1.5 trillion after breaching briefly \$2 trillion in May. This remarkable expansion in value from zero in 2009 has not gone unnoticed by the general public and has been observed with astonishment by practitioners of legacy finance, but it still represents 
less than one percent of global wealth\cite{suisse2020global}.

Many attempts exist within the technical constrains of the trust- and permission-less framework governing cryptofinance to translate existing and to construct novel financial products and structures.  This area is called Decentralized Finance (DeFi). It covers exchanging, lending, and tokenizing\footnote{In the rest of the chapter the term `token' will be used loosely. We will not be restricted to the UK's Financial Conduct Authority (FCA) definition: \url{www.fca.org.uk/firms/cryptoassets}. } of assets and has recently in a variety of metrics, e.g. Total Value Locked (TVL), transacted volume, products available and number of participants, increased dramatically. In the past year, there has been a jump by a factor of 60   according to DeFi Pulse of the ``collateral for loans, trades and other transactions in DeFi applications" \cite{disrupt_finance_2021}, and according to the data aggregator DeFi Llama \footnote{DefiLlama - DeFi dashboard, \url{www.defillama.com}.
} the TVL, one of the key measures used to estimate market size, in DeFi  first  breached $\$ 150$ billion  in May 2021.  The constantly changing amounts and rankings for popular DeFi products and platforms can be found on the public dashboards of DeFi Llama.

Crypto-optimists regard DeFi as a way to improve efficiency, transparency, interoperability, accessibility, and inclusivity beyond what has been achieved in traditional finance, even if  unlimited transparency and unimpeded access do not necessarily entail social and economic benefits  -  see for example an article by Dennett {\it et al.}\cite{dennett_2015_our}.

 For the purpose of this chapter yield based investments are defined as the combination of a positive return accumulating in a regular almost clock-like fashion  with a short position in an out-of-the-money option corresponding to tail risk. An example from legacy finance would be a credit-risky bond which combines interest accrual with the downside risk of  the issuer declaring bankruptcy and defaulting on the obligation to pay interest and principal.  Due to the heightened volatility of cryptocurrencies yield based strategies might seem like trying `to pick up pennies in front of a steam roller', but with rigorous portfolio diversification the idiosyncratic risk associated with tail events often becomes manageable.
 
In the derivatives ranking by notional of legacy finance fixed income dominates, see  the statistical release for 2020 of the Bank of International Settlement\cite{BIS_otc_derivatives_december_2020},
but this does not extend to gross or net exposure, since volatilities of interest rates are often  smaller by an order of magnitude than volatilities of other asset classes.  This already suggests a challenge for DeFi fixed income products, since crypto collateral is normally linked to the notional amount and not exposure, whereas  margin payments between counterparties via exchanges, clearing houses  or credit support annexes (CSAs) in legacy finance are mostly based on the more dynamic and complex to determine  net exposure\footnote{Attempts in DeFi to develop zero knowledge  dynamic credit risk monitoring and margining systems, e.g. X-Margin,  have persuaded some VC investors, but have gained up to now limited traction.}.

Positions on a blockchain updated by Proof-of-Work (PoW), most prominently Bitcoin, earn zero return similar to holdings in cash\footnote{ In conventional interest rate modelling cash is usually ignored, since a central bank backed riskless short rate exists, but for cryptocurrencies ignoring positions directly held on the blockchain  is not feasible.}. This implies a vanishing short rate, which clashes with  `conventional' interest rate models, which take a non-zero short rate to be a `primitive', i.e. a necessary building block for  term structure, discount rates and a requirement to flexibly model pay-off structures. In a paper by Brody {\it et al.}\cite{brody2020theory} it was shown how to overcome this difficulty.

For  blockchains with Proof-of-Stake (PoS)   updating, e.g. Ethereum is in the process of switching from PoW to PoS, staked tokens  have a non-zero yield, which relates a largely deterministic return to the amount committed. In addition, validators are able to exploit maximal/miner extractable value (MEV) linked to the ability to select and order transactions from the Mempool for blockchain inclusion, but  also incur in the process obligations.
Non-staked tokens  can participate in DeFi or  gain convenience yield from immediate accessibility, which can   for example pay for imminent  blockchain fees. The risk-adjusted benefits for the marginal investor in these three forms of token usage have to match to avoid runaway effects. 
The relevant return data for staked  Ethereum is presented in section 2.
 
Besides conventional fixed income products there are other areas of DeFi where  
yield-like returns exist. Decentralised exchanges (DEXs) provide such an opportunity to liquidity providers (LPs), who lock-up token pairs in pools to allow liquidity takers (LTs) to switch between them for a fee. The LPs garner the fees but are short an option associated with exchange rate fluctuations. This will be discussed in section 3.

DEXs are alternative liquidity sources with universal   access. A   natural 
extension
of the crypto-anarchist ideals set forth in the cypherpunk movement of the 90s and crystallised in Bitcoin by Satoshi Nakamoto. By Removing the middleman, as well as any form of control, DEXs are furthermore a response to failures of legacy finance. 
On centralised exchanges (CEXs) users are required to transfer custody of their assets to a third party, which entails risk. The temporary freezing of assets as well as the collapse of exchanges, e.g. Mt Gox of Japan in 2014, has shown that this risk is non-negligible.
DEXs are increasingly coming under official scrutiny,   
as they circumvent money laundering rules,  
and mix assets without regard for regulations.

Another area of interest is modelling.    For the cryptocurrency interest-rate derivatives market to bloom, pricing, replication and hedging need to function smoothly. Models are required that are convincing and common to market participants. This is a necessary but not sufficient condition for the establishment of     a capital efficient  crypto fixed income market. A way to evaluate a simplified version of these loans without the detailed examination of the liquidation mechanisms, which differ from platform to platform, is given in section 4.

A proposal for a crypto product linking collateral not to notional but exposure is presented in section 5. A natural candidate is a modified version of a cross currency swap traditionally used by banks to exchange collateral and interest payments in different currencies without taking directional foreign exchange risk. 
Section 6 covers yields of short dated loans on CEXs.
  
Perpetuals swaps traded on centralised exchanges enable speculators  to leverage their capital close to a stratospheric hundred times.  Speculators are willing to pay a premium over spot to access leverage motivated by the chance to invest in possibly highly lucrative momentum strategies, whereas liquidity providers get a return associated with the size of the premium but have to take on unpopular positions with  drift and jump risk.  
The basis of perpetual and finite duration futures is the topic of section 7.  
The interest rate implied by crypto-options is the topic of section 8.
A conclusion rounds off the chapter. 
 
\section{    Protocol Defined Return (PDR) and Ethereum 2.0 }
\noindent 
In this section information about the evolving Ethereum 2.0 protocol and the associated Beacon Chain is presented. The Ethereum 1 blockchain is currently separate from the Beaconchain to reduce risk in the transition. The new protocol is not finalized and evolves with the EIPs (Ethereum Improvement Proposals \footnote{https://eips.ethereum.org}). The novelty of the protocol and the complexity of the switch from PoW to PoS involving multiple stages has led to a lack of data for staking returns in the academic literature.   Before we present the data, which covers the period from December 2020 (Launch  of the Beacon Chain) to September 2021, a rough introduction to some relevant parts of staking as it exists in June 2021. To participate in staking at least $32$ ETH has to be pledged for the whole trial period. The volume weighted staking return as used by Attestant \footnote{https://www.attestant.io/metrics/aor/}, $R_{stake}$ between time $t-1$ and  $t$ for  an eligible validator is given by

$$ R_{stake} = 365 \times \left( \frac{V_t}{V_{t-1}} - 1 \right), $$

\noindent where $V_{t-1}$ is the validator balance at time $t-1$, and $V_{t}$ is the validator balance at time $t$. The validator balance is taken always at $00:00$ UTC. 
An eligible validator is a validator that is in the `Active' state at time $t-1$ and   
remains continuously in the `Active' state until, and including, time $t$. The validator must also maintain a  
minimal balance of 32 ETH during the entire period. Balance changes in figure 1 are obtained from all validators active within the previous 24h hours\footnote{Data provided by \url{www.attestant.io}}. The Ethereum 2.0 upgrade has incorporated a further range of interesting features. For the more  technical aspects we refer readers to the voluminous Ethereum 2.0 documentation. This includes the work a validator has to perform  and how, as a result, the  balance $V_{t-1}$ is converted into $V_t$.  

Some pertinent features necessary for understanding   figure 1 are that stakers can neither at this stage withdraw funds back to Ethereum 1 nor can they switch funds to a different validator to increase their yield. As a result, the ability  to optimise allocations between validators is restricted.   
 In addition, there is a ceiling on individual validator balances 
 eligible  to receive returns, and initially high returns are whittled down as the surplus  unable to generate returns increases. However it must be noted that when Ethereum 2 is fully operational validators will receive fees from transactions as well as block rewards increasing returns for staked assets.

\begin{figure}[h]
\includegraphics[width=\columnwidth]{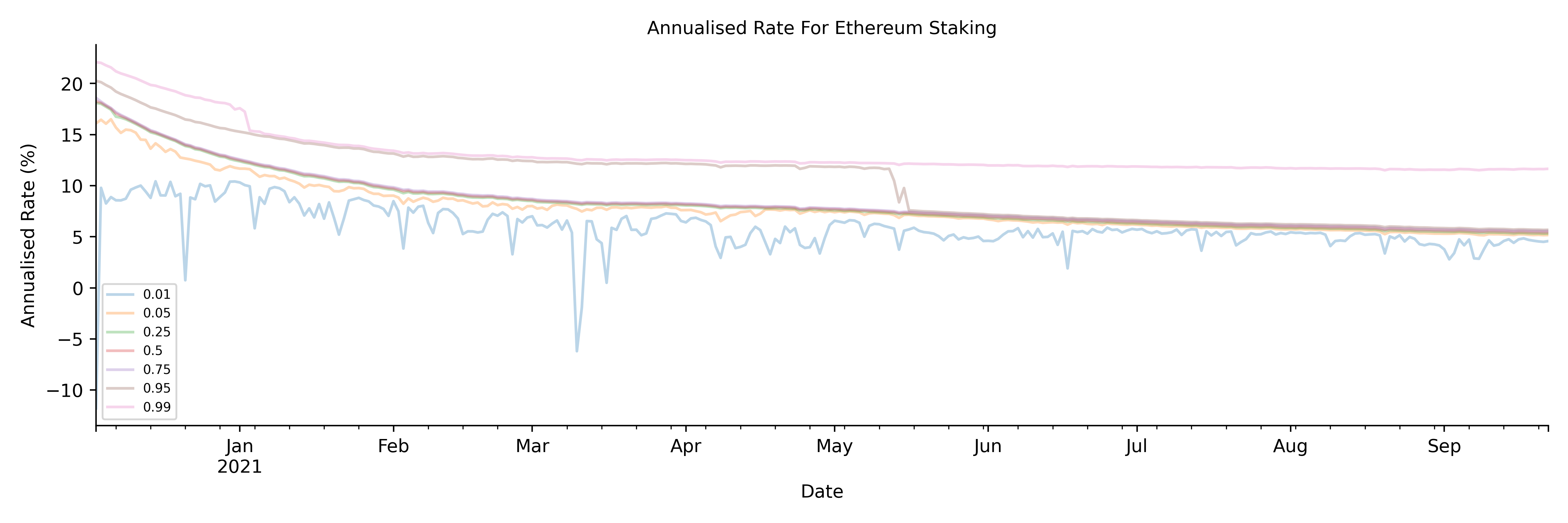}
\caption{Staking rewards  for all validators active in each 24h period; shown are returns in percent of the 1\textsuperscript{st}, 5\textsuperscript{th}, 25\textsuperscript{th}, 50\textsuperscript{th}, 75\textsuperscript{th}, 95\textsuperscript{th} and 99\textsuperscript{th} percentile. Source: \url{www.attestant.io} } 
\end{figure}

\noindent  
Features of the chart are:
First, returns started high, are slowly declining and converging, i.e. the spread between different percentiles is decreasing. 
Second, the erratic returns of the  1\textsuperscript{st} percentile, which on occasion have been negative, are due to a failure to adequately fulfil the required tasks. Negative returns can be caused by `Slashing', when incorrect information is validated, or to a lesser extent   from inaction. Punishment is increased with the amount of network disruption. For example, all validators, who failed to validate, are slashed jointly using the following equation: $Slash ~ Cost = 3 \times N_{\%}$. As a consequence, a pool controlling $1/3$ of the network could have a $100\%$ loss. 
Third, there are also those validators that perform exceptionally well, i.e. they  perform a special validator function. This allows them to choose a block for which they are rewarded in ETH, which is reminiscent of Bitcoin mining. However,  
unlike in Bitcoin, these special rewards are based solely on luck  
and not influenced by the computational capacity of the validators.  
 Except for this random element, returns  are more predictable than in PoW and directly follow from the number of active  validators, which is  a known quantity. 
Fourth, the `step functions' observable in the return of the highest two percentiles could either be due to protocol changes or discrete effects linked to the number of validators winning  top tier rewards, usually given to block proposers in the ETH committee system. 
As the overall number of participants grows, the chance of obtaining these slots diminishes. 
To determine if protocol changes or marginal validator returns are responsible for jumps one only needs to consider a more fine-grained percentile graph. 

Dedicated standalone entities  provide both  bespoke and  non-pooled services, where funds are segregated, not mixed in validators, and security is ensured.
Exchanges  also offer for a fee staking services to enhance yield on positions under custody.  

Although Ethereum  2.0 is perhaps the highest profile staking protocol, it is not the first to adopt  PoS.  The main other staking cryptocurrencies are Tezos, which calls staking  
`baking', Algorand and Cardano. The protocols differ, but all rely on dedicated
`collateral' to perform the network verification function. 

\section{Automated Market Makers and Yield}
\noindent
DeFi has expanded into a number of different directions. In  subsection A,  Automated Market Makers (AMM), a core component of decentralised finance, are introduced, and in subsection B the financial mechanics of AMMs are discussed. In the final subsection, some challenges facing AMMs and the yield  of an exemplary AMM is  presented. 

\subsection{Introduction to Automated Market Makers}
\noindent
Arguably the biggest growth  has been in AMMs
facilitating token exchanges,  which are in competition with centralised exchanges (CEX). Big participants, or maybe they should already be called incumbents despite the novelty of the field, are for example Uniswap, Balancer, Curve, Pancakeswap,  and Sushiswap. On these DEXs liquidity providers, which earn yield in the form of a stream of commissions but are exposed to jump risk in the token exchange rate, are matched to liquidity takers wanting to exchange one token for another.

AMMs are substantial. Uniswap,  which is in its third major incarnation, alone has executed as of June 30\textsuperscript{th} 2021  trades  worth almost $\$300$ Billion, supports $40,000+$ token pairs of different sizes and has combined liquidity of over $\$3$ Billion according to the Uniswap website. These numbers are impressive for any exchange, but especially for an originally one person project\footnote{ `A short history of Uniswap' can be found at  \url{https://uniswap.org/blog/uniswap-history/}.} started in 2017 on a shoestring. It then moved through various stages and was deployed on the Ethereum Mainnet in November 2018. It is based on open-source code, which can and has been forked and adapted by competitors such as Sushiswap.   

Uniswap and other AMMs are not directly comparable to traditional exchanges, since their legal status is both contentious and recognisably different. A simplistic way to highlight the difference  is by pointing out the distinction between the verb `to exchange' and the noun `the exchange'.    The Uniswap AMM should be viewed as a software protocol deployed on the Ethereum blockchain built and supported by an entity called Uniswap with a peculiar institutional structure controlled by the UNI governance token  held with variable lock-ups by users, team, investors  and advisors with a market capitalisation of just under $\$$10 Billion as of June 2021. 
\subsection{ The financial mechanics of Automated Market Makers}
\noindent
This section serves as a limited introduction to the  financial mechanics  of AMMs  as employed by Uniswap and similar protocols; details of these mechanics are subject to change. For full details of the evolving smart contract implementations see the websites of the providers, which also explain how  specific tokens, e.g. UNI of Uniswap, are created to differentiate protocols from competitors, to attract liquidity and to deal with governance issues including the determination of contract specifications.  
One major challenge faced when translating the functionality associated with limit-order based exchanges, which offer a plethora of order types and differentiate between market participants, 
 to the distributed Ethereum Virtual Machine was to reduce memory and computational requirements.  Many AMMs including Uniswap use a simplified market maker model associated  with the acronym CPMM (`Constant Product Market Maker').

\noindent

In this type of  market  each token pair is called a pool and divided into two types of participants. These are liquidity providers (LPs) and liquidity takers (LTs). The smart contract employs a simple deterministic formula to incorporate the two types of actions each market participant is able to perform after a pool is created.  The key formula is $xy=k$. $x$ and $y$ are respectively the two tokens which form a particular pool. The ratio $x/y$ can be viewed as the exchange rate associated with the pool. Details can be found in  a paper by Y. Zhang {\it et al.} \cite{zhang2018formal} and in the whitepapers and other materials of the various AMM platforms.
The pool is stocked by liquidity providers (LPs), who can add and subtract liquidity. This means LPs provide or withdraw supplies of tokens at predefined ratios. Subtleties involving the submission of a combination of tokens which do not exactly match the pool ratio are ignored.  
The LTs also have two ways to interact with the protocol. LTs can either send an amount of the first token in the pool to receive an amount of the second token, or they can do the reverse. The amounts received are predetermined by the pool composition and incorporate a fee proportional to the trade size to incentive LPs to commit liquidity. As these four interactions with the smart contract governing the pool occur the composition of the pool is dynamically adjusted. 

Arbitrageurs observe  a host of DEXs and  CEXs and ideally keep  prices aligned within the limits set by transaction and other frictional costs. As a pool gets larger, i.e. LPs add liquidity, the market impact of an LT order of fixed size is reduced, whereas a reduction of tokens in the pool increases market impact. The reverse is true for LTs. If an LT wants to switch a large amount, then the execution price moves further. This is a form of price impact or slippage. 

The fee charged in a pool can be adjusted as part of the governance procedure involving UNI tokens. The current fee on Uniswap V3 is tiered,  $0.05\%$,  $0.3\%$, or $1\%$, and depends on the token pair chosen. Pools with higher volatility are riskier and normally require a higher fee. Various costs due to interaction with the blockchain, e.g. gas in the case of Ethereum, are the responsibility of LPs and LTs and can lead to inertia. 
DEXs  are dynamic and constantly tinker with their business model. In Uniswap V3, the newest version of Uniswap launched in May 2021, liquidity providers can concentrate liquidity  in particular exchange rate ranges, which is of particular benefit for stable coin token pairs with smaller volatilities and also lower fees.

LPs lock-up their assets when they enter a pool. On exit LPs get a proportion of the accumulated fee commensurate with their relative stake in the pool plus an appropriate proportion of the liquidity pool itself.  
The fee creates an incentive and is similar to a yield, but it is tempered by the risk associated with being short volatility in the exchange rate of the underlying token pair. This risk is  slightly incongruously called `impermanent loss'.  It is the relative loss due to the change in the pool composition from the LP's entry and exit point. In other words, if the investor had retained the initial position ignoring the accumulated fee,  then this would have been more valuable than getting the final mix of tokens, which reflects the change in the pool exchange rate.  Examples are provided in Y. Zhang {\it et al.} \cite{zhang2018formal}. Besides   relative losses  one can also consider numeraire dependent absolute value changes of the pool portfolio.   
This comparison  can show both gains and losses and is depicted in figure 2, which  shows  for USD as the numeraire the   
cumulative absolute impermanent  gain or loss   
 for the ETH-USDT pool. 
Absolute value changes are of importance for the overwhelming majority of investors, who have either reporting or financial obligations in a specific currency.  
\begin{figure}[h]
\includegraphics[width=\columnwidth]{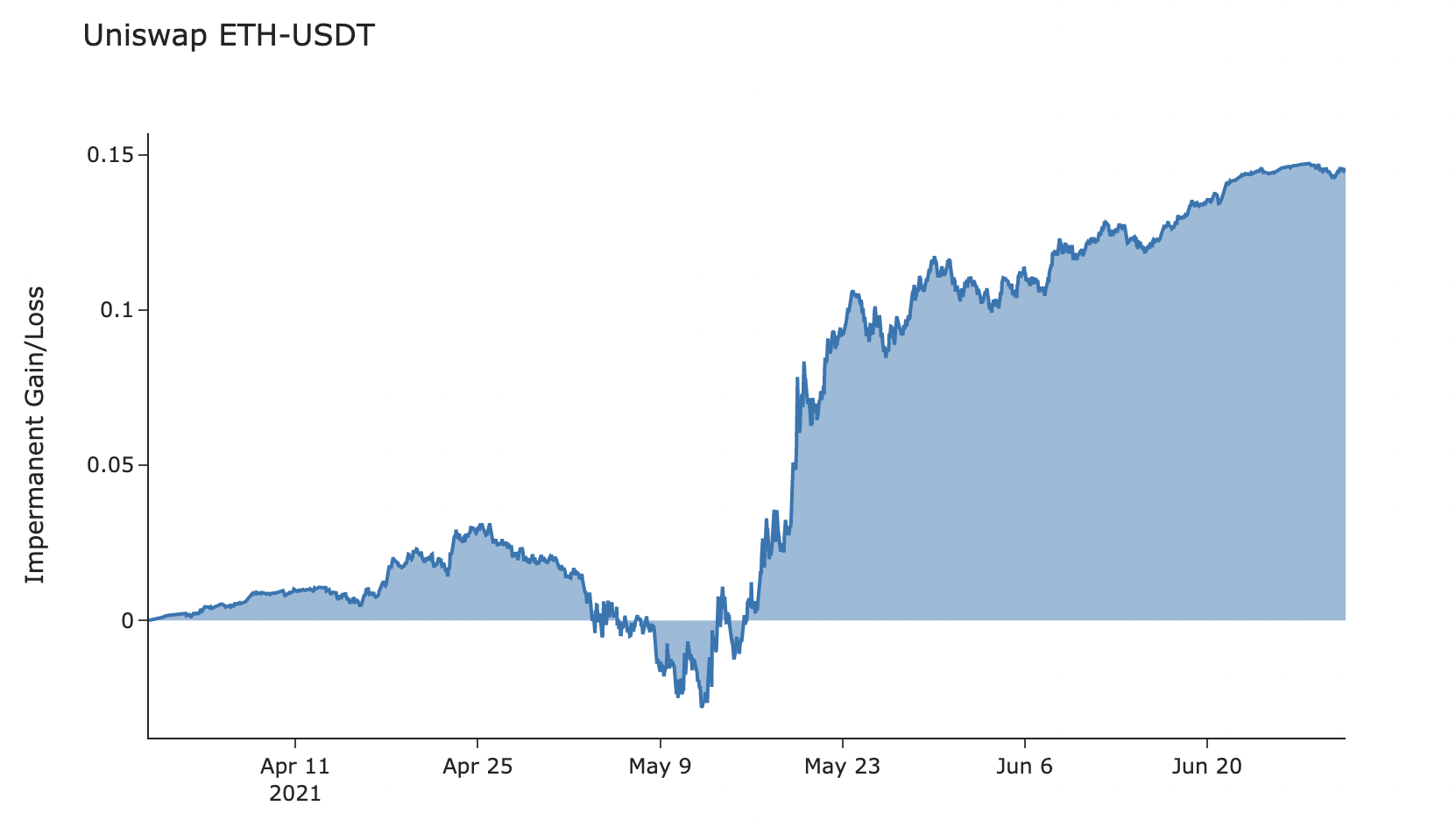}
\caption{Cumulative Impermanent Gain/Loss for the  Uniswap ETH-USDT token pool as fraction of initial notional with USD as the numeraire, Data Source: www.intotheblock.com}
\end{figure}

\begin{figure}[h]
\includegraphics[width=\columnwidth]{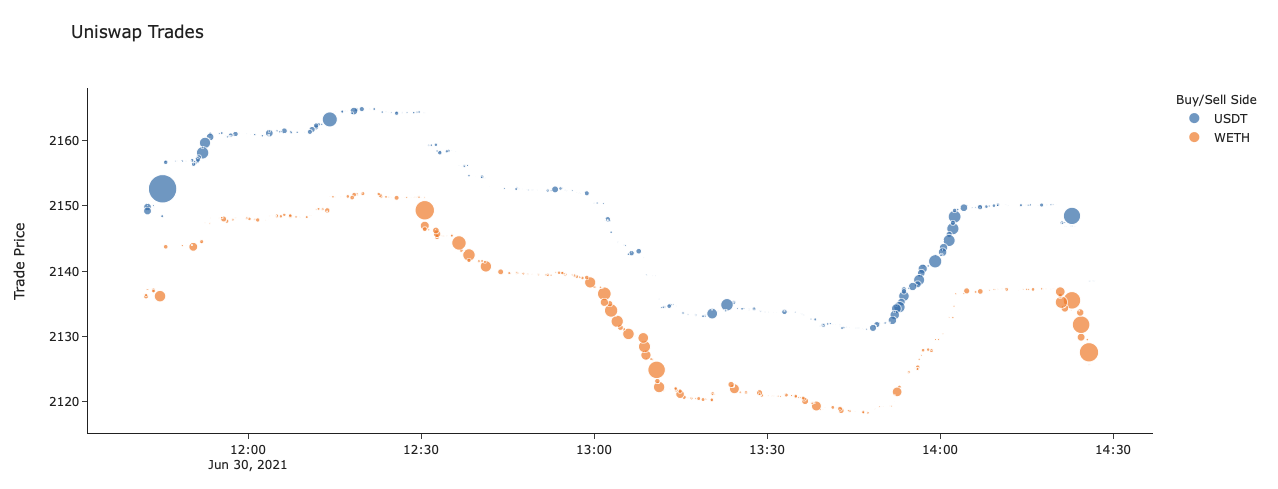}
\caption{Two hours in the life of the Uniswap WETH-USDT Token Pool, Source: pools.fyi}
\end{figure}
 
In figure 3, the price dynamic  of an Uniswap token pool controlled by an AMM is captured beautifully. In this case it is the token pair of wrapped ETH (WETH, but subsequently written as ETH) and the stable coin Tether(USDT). What would normally correspond to the bid-offer spread on CEXs is dependent on the $0.3\%$ fee and the pool size. The bid-offer spread on DEXs increases with the size of the fee and decreases with the size of the pool. 
As the price drops in a conventional exchange `bids are hit', but on an AMM tokens are exchanged (orange circles: ETH is exchanged for USDT) such  that the price drops.
Trade size is depicted by the circle area, which seems to expand when prices move. There is asymmetry in up- and down-trending markets, i.e. sell-side trades  are more prominent in a downward move and buy-side trades more prominent in an upward move. These concentrated larger trades with the commensurately  larger slippage can be profitable when markets trend and are likely to have originated with arbitrageurs exploiting prior price moves on CEXs. From this graph alone one cannot tell with certainty if the Uniswap DEX leads or lags other DEXs or CEXs. The hypothesis would nevertheless be that the liquidity pattern observed is suggestive of this DEX being a follower. This is bolstered by the fact that CEXs can react faster to market news than DEXs dependent on sluggish blockchain updating times.  

\subsection{Challenges of  Automated Market Makers}
\noindent
In this subsection challenges facing AMMs and  yield data is presented. The competition between pools exists in multiple dimensions and is a fascinating example of game theory in action. Yield-enhancement subsidies to increase the competitiveness of particular protocols using platform specific tokens  lead to possible distortions. Benefits for LPs can be detrimental for LTs and vice versa. Balancing short-term and long-term greed can be particularly  difficult and might not have stable solutions in such a transient and transparent environment.

Let us briefly describe the case of a LP participating in an idealized pool for the duration $T$. If for example the exchange rate follows approximately a random walk with volatility $\sigma$ per unit of time  then the `impermanent risk' is expected to be proportional to $\sigma\sqrt{T}$,
whereas  the  average fee of $\alpha$ per unit of time should accumulate linearly at the rate $\alpha T$, and should eventually dominate. The amount gained on average per unit of time in the large $T$ limit is  $\lim_{T\rightarrow \infty} (\alpha T -\sigma \sqrt{T})/T = \alpha$. 

A way to compare various pools is to look at the associated Sharpe ratios, which are $(\mu_i-r)/\sigma_i$, where $r$ is the riskless rate, $\mu_i$ is the return and $\sigma_i$ the volatility of the $i$\textsuperscript{th}-pool. The riskless rate in many major fiat/numeraire currencies  is currently approximately zero, and from the time series of returns the Sharpe ratios can be computed. Ignoring correlation and instead assuming the time series of the pools are approximately independent, which is an extreme idealisation, one  could rank the pools in a simple way. Portfolio optimisation using the Kelly criterion can be applied\cite{ kelly_1956_a,lv2009application,lv2010implication} to optimize investments in pools taking the covariance of returns into account.

AMMs are accessed through the  blockchain, and preferential access leads to profits\cite{zhou2020high}. 
Miners can select and reorder Mempool submissions, e.g. front- and back-running orders, to create MEVs  - see Daian {\it et al.}\cite{daian2019flash} for an early discussion. 
This has created a hyper-profitable cottage industry together with service providers such as flasbots.net, who consider what is arguably
a vice, i.e. a (possibly unavoidable) friction cost for investors,  as a 
 profit-generating virtue.

\section{Over-Collateralised Cryptoloans}
\noindent
This section on cryptoloans divides into three subsections. In  subsection A cryptoloans are introduced and market data is presented, in the subsection B a toy model is analysed, and in the last subsection liquidation mechanisms are described. 
\subsection{What are Cryptoloans?} 
\noindent
Cryptoloans are a form of over-collateralised lending managed by a smart contract. Potential lenders send tokens, often in the form of stable coins, to a lending pool, whereas the borrowers deposit collateral often in the form of Ether and in return gain access to a fixed amount of stable coins. Due to the trustless nature of the exchanges the value of the collateral is around $150\%$ of the borrowed amount. There are also off-blockchain providers where  interest rates are tiered with the level of initial collateral. If the value of the collateral falls below a certain threshold compared to the borrowed amount a liquidation procedure ensues to protect the lenders. Liquidation normally involves a penalty and can be in the form of a fixed time unwind auction or based on an oracle specified price. For an extensive analysis see a recent paper by  Kaihua Qin and co-authors mainly from Imperial College\cite{qin2021empirical}.  Significant deleveraging occurs when token prices jump  dramatically, as has happened on the 19\textsuperscript{th} of May 2021. 

The attempt to automatically re-margin or faithfully unwind  the initially over-collateralized loan  through a smart contract appeals, but it is technically cumbersome and as a consequence involves operational risk,  and more importantly is costly in   on-chain computation  \& data.

The interest rate borrowers pay depends on the volatility of the token pair and the amount of collateral required, but also on the utilization of the lending pool. The interest rate model used differs from platform to platform. On Aave it has the form of a hockey stick.  For a lending pool utilization of up to $80\%$ the interest rate increases slowly, but above this utilization rate it increases at a steeper slope. This is to ensure that lenders of tokens in demand benefit from scarcity. Other platforms, e.g. MakerDAO, Compound and Curve, provide variations of the theme.   
More details of the ever-changing protocols can be found on the respective websites. The constantly shifting rankings for loan platforms as well as other DeFi products can be found on the website DeFiLlama.com.
Figure 4 shows the volatile borrowing and lending rates in June 2021 for USDT  on Compound  with ETH as  collateral. The rates are a multiple of what is available for USD in legacy finance and an indication of market segmentation. 
\begin{figure}[h]
\includegraphics[width=\columnwidth]{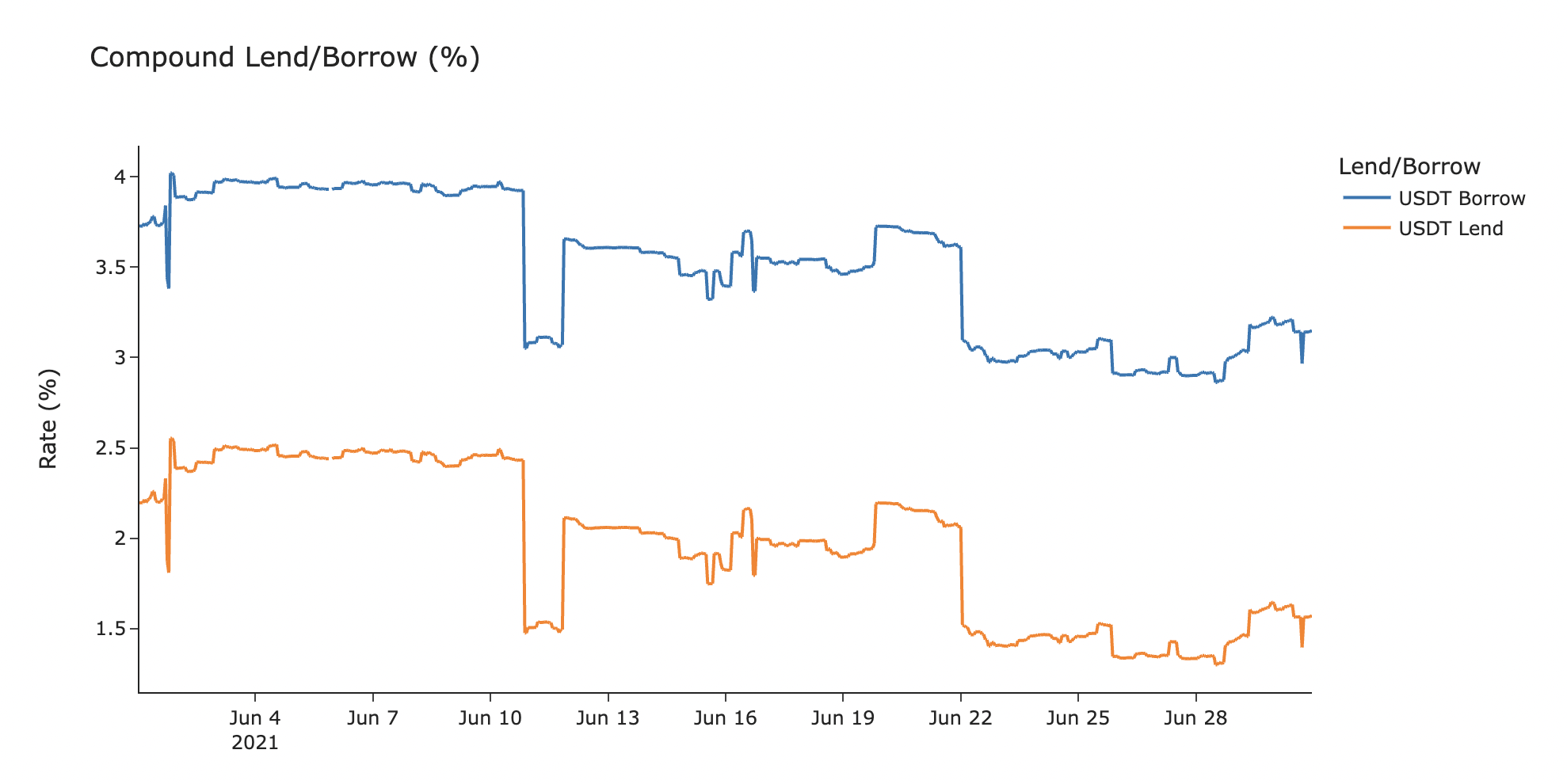}
\caption{ Annualised Interest Rate for borrowing and lending of USDT on Compound with ETH as collateral, Source: Compound}
\end{figure}

\subsection{Pricing and Modelling of Cryptoloans}
Cryptoloans need to be modelled and priced.  We next present a simplified evaluation based on an option introduced by Margrabe\cite{margrabe1978value} and consistent with Black-Scholes-Merton option theory, where one party has the right to choose between two risky assets at maturity. 
The model has its limitations, since the underlying price process is assumed to be a geometric Brownian Motion, but it  can provide a starting point for further investigations.  Collateralised options can be priced for more general price processes - including jumps or stochastic volatility. 
The main weakness 
of the model is the neglect of the liquidation process. This will be partly rectified
in the next subsection.

One can view a collateralised  loan in the static hedge case as an option on the minimum of two assets, since the token amount used to over-collateralize the loan can fall below the value of the loan notional plus interest at maturity. The value of the option in the Black-Scholes-Merton model  is dependent only on the interest of the two tokens, their correlation  and their respective implied volatilities. 

The payout for the borrower is $Max[A,B]$, where $A$ is the collateral in the form of token $\alpha$ and $B$ the amount of token $\beta$ to be paid at maturity time $T$. The payout for the lender is ${\rm Min}[A,B]$, and the sum of the pay-outs is the $A+B$. The value of such an option at time $t=0$ is:
\begin{eqnarray}
Max[A,B]&=&B+Max[0,A-B]\nonumber\\
&=&B+e^{-r_{\alpha} T}A N(d_1) + e^{-r_{\beta} T} BN(d_2) ,\nonumber
\end{eqnarray}
where $N(\cdot)$ is the standard normal cumulative distribution function, and
\begin{eqnarray}
d_1&=& \frac{\log(A/B)+r_{\beta}-r_{\alpha}+\frac{1}{2} \sigma^2 T}{\sigma\sqrt{T}},\nonumber\\
d_2&=&d_1-\sigma\sqrt{T},\nonumber\\
\sigma^2&=&\sigma_{\alpha}^2-2\sigma_{\alpha}\sigma_{\beta}\rho_{\alpha,\beta}+\sigma_{\beta}^2,\nonumber
\end{eqnarray}
with $\sigma_{\alpha}$ and $\sigma_{\beta}$ the volatilities of tokens $\alpha$ and $\beta$ respectively, and $\rho_{\alpha,\beta}$ the correlation between the two tokens. The interest rates $r_{\alpha}$ and $r_{\beta}$ are the  interest rates of the two tokens. All the prices, volatilities and interest rates are given with respect to the numeraire, e.g. the US dollar. If one of the two assets is the numeraire itself (or an asset with zero volatility \& drift with respect to the numeraire like an eternally tethered  stablecoin),  then the formula  simplifies. For example for $A$, as the notional amount denominated in the numeraire,  it becomes
\begin{eqnarray}
B+Max[0,A-B]=B+ AN(d)+ e^{-r_{\beta} T}B N(d-\sigma_{\beta}\sqrt{T}),\nonumber
\end{eqnarray}
with
\begin{eqnarray}
d= \frac{\log(A/B)+r_{\beta}+\frac{1}{2} \sigma_{\beta}^2 T}{\sigma_{\beta}\sqrt{T}}.\nonumber
\end{eqnarray}
The European and American versions of this option have equal values, since early exercise is under the artificial conditions considered, i.e. no dividends and the underlying process is a geometric Brownian motion, never advantageous. The existence of multiple collateral tokens, as exist in some protocols (Balancer has up to 8 for some markets), complicates the calculation of maximum and minimum values. A solution for prices limited to geometric Brownian motions was discussed in `Options on the maximum or the minimum of several assets' by Johnson\cite{johnson1987options}. 

\subsection{Liquidation of Cryptoloans} 
\noindent
By ignoring the  liquidation procedure  a simple but insufficient formula was derived.  As a next step, one can incorporate a pared down version of a liquidation penalty function,  which awards a fixed amount, i.e. a fraction of the notional, to the lender, if some collaterlization limit is reached.  This could be modeled as a one-touch option\footnote{For the value of one-touch options in the Black-Scholes-Merton framework and beyond see chapter 12 of the comprehensive book by A. Lipton \cite{lipton2001mathematical}.}. The liquidation penalty on the lending platform Compound is   8\%, while on Aave it is 5-15\%. 
The value of a loan is in the simplest penalty case, where margins cannot be replenished,  the combination of two options, which can still be priced directly. Various subtleties inherent in the actual liquidation process involving for example time delays 
or the existence of a centralised recovery fund\cite{qin2021empirical}
inhibit accurate option replication and add non-negligible noise.

The ability to disentangle loans in terms of a combination of options provides a useful way to build a bridge to the  nascent crypto options market. It offers a way to  price and hedge crypto options  and allows liquidity of the loan market to leach into the less liquid options market. 

A way to enhance the utilisation of collateral would be to allow the lending contract to re-lend part of the collateral, i.e. a form of recycling, in a separate loan. The advantage would be a more efficient use of capital, whereas the disadvantage would be 
the increased risk of losses. As collateral chains get longer the risk, if  the various exchange rates are independent, would be approximately additive, but  if it involves triangles of currencies, i.e. from A to B to C and back to A, then the risk would stay in a tighter range.

\section{Efficient Use of Collateral to Enable Leverage and enhance Yield}
\noindent
One of the issues that has plagued  crytpo-finance is the difficulty to generate leverage.
In this section we propose a structured product as a prototype solution. It is based on a cross
currency swap and separates credit from foreign exchange risk. These swaps are widely used by for example  Japanese banks\cite{boj2021}, who fund themselves in their home currency at a stable low rate,
but lend internationally in dollars. 
Cross currency swaps are a tailor-made solution, because they allow  Japanese banks to fund positions in a foreign currency
without the associated foreign exchange risk.
 Credit risk Japanese banks can manage through   
the construction of a diversified portfolio of loans, whereas foreign exchange risk is concentrated and would otherwise 
necessitate a hedge.

The contract starts with a notional exchange denominated in two currencies at an agreed exchange rate
normally close to the current spot rate and it ends at the termination date with a reversal of the notional exchange at the same initial exchange rate. At regular intervals during the lifetime of the swap cash flows linked to either fixed or floating rates in the two currencies
plus a potential spread  are exchanged. 

Next we describe how this can be translated into the trustless cryptocurrency framework\footnote{MK2 Finance in collaboration with others is developing such a contract.}, and  why it is beneficial. The first and last step would be identical, i.e. the initial and final exchange of notional via an intermediary smart contract, allowing both sides to reuse the notional. To avoid losses due to currency movements each side is required to deposit an initial margin  into the  smart contract.
An oracle monitors the exchange rate and observes if the margin of either side falls below some critical threshold. If this occurs and the margin is not replenished the contract is prematurely terminated. Each side keeps the exchanged notional amount and the party, whose margin limit has not been breached, acquires  enough of the margin of the other party to be made whole.  
A version of this contract without an oracle, but with  either side able to prematurely terminate the contract for a fixed  fee is also possible. 

Details about buffers and various other payments, which take into account supply and demand, as well penalties for early liquidation require careful consideration. One should note that an unexpected jump in the exchange rate can exhaust the margin buffer. Therefore buffers should take into account the volatility of the exchange rate. Asymmetry of possible movements could as well as supply and demand lead to the initial and final exchange rates or the margins to differ by a market determined factor. The importance for both sides to have access to the notional minus a margin enhances symmetry, currently not available in crypto-loan deals, and will fuel reuse of the notional. 
 
The buffer corresponds to a hair-cut and might reasonably increase with the square root of the duration of the arrangement. If the buffer corresponds to $x\%$, then the total leverage  gained by each side, assuming maximal recycling of the notional, would be slightly under $1/(x\%)$.
This  turbo-charges crypto-finance but  entails risk, since the underlying historical volatilities guiding the choice of $x\%$ might change spontaneously across the board leading to wild deleveraging. Even with the current over-collateralized contracts DeFi TLV dropped by a third in the recent period of volatility in May 2021, see the dashboards of DeFi Llama. 
\section{Short Dated Cryptoloans on Centralised Exchanges}
\noindent
In the subsequent short sections we switch from DeFi to Centralised Finance (CeFi). We start with the presentation of data of different short dated loans from the crypto-exchange Bitfinex.
\begin{figure}[h]

\includegraphics[width=\columnwidth]{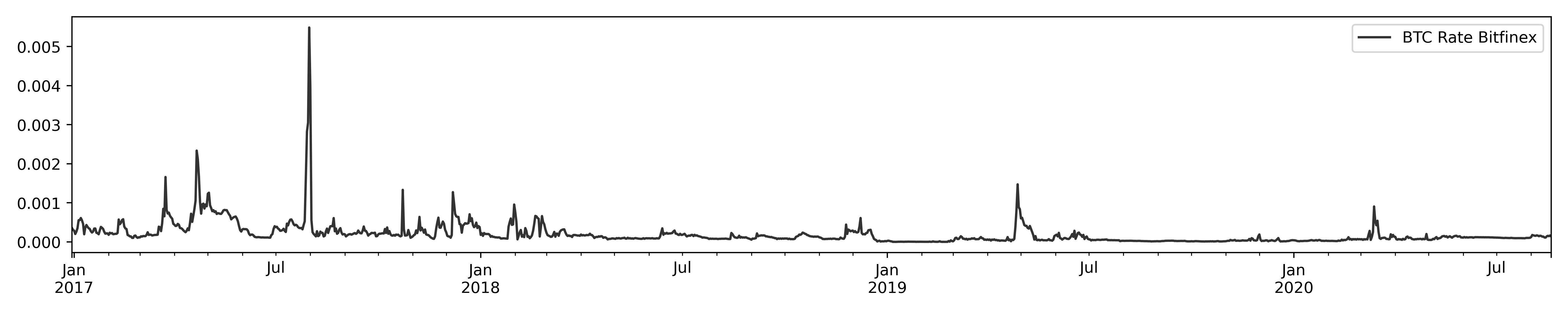}

\includegraphics[width=\columnwidth]{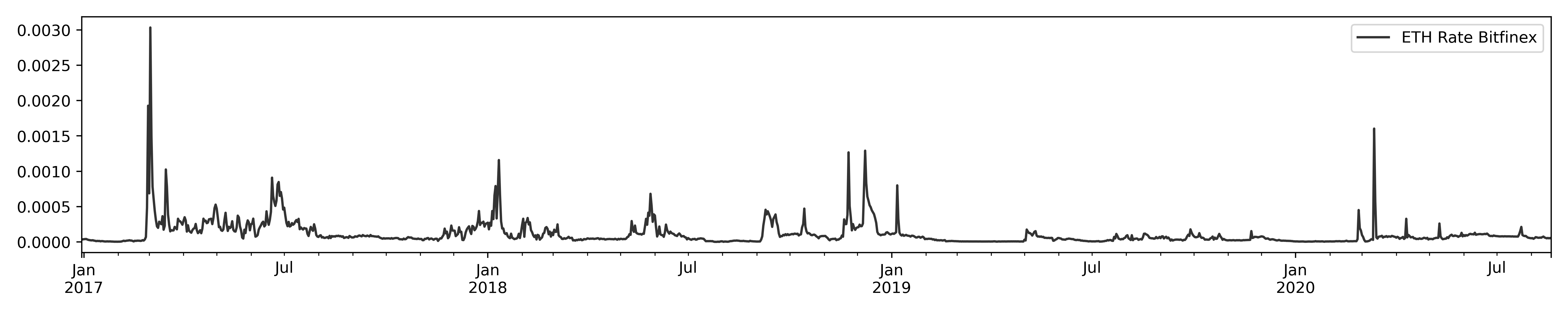}

\includegraphics[width=\columnwidth]{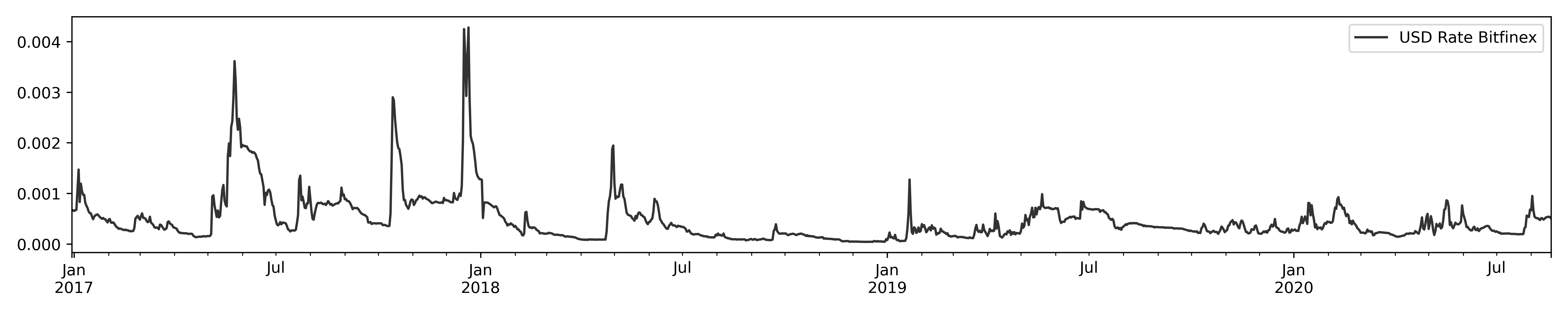}
\caption{ Weekly rolling average of the interest rate in \% per 365 days on Bitfinex  for short dated BTC, ETH, and USD  loans 
(for details: \url{www.docs.bitfinex/v1/reference\#rest-public-fundingbook})}
\end{figure}
The three graphs of figure 5  show the lending rate per day for BTC, ETH and USD loans of two days, which is the most liquid duration. A limit order book is used to match borrowers and lenders. The market has existed comparatively long and provided an early opportunity to establish leveraged positions in cryptocurrencies.

\section{ The Basis of Perpetual and Finite Duration  Futures}
\noindent
In this section, we study different types of CEX futures. Accessibility of leverage differs in the futures and spot   markets. 
In subsection A there is a discussion of the history of perpetuals and some related market data and in the subsequent subsection the same is done for finite duration futures.

\subsection{The Basis of Perpetual Futures} 
\noindent
In the early 90s Shiller\cite{shiller_1993_measuring} introduced a novel way for pricing perpetual claims on indices. Together with Case he  introduced a real estate based index, which after multiple re-incarnations turned into the  `S\&P CoreLogic Case-Shiller U.S. National Home Price Index'. Futures and options written on this index are listed on the Chicago Mercantile Exchange(CME). Undated gold futures traded already  in Hong Kong  in the 1980s and can be viewed as precursors of perpetual futures\cite{gehr1988undated}. 

Starting from 2016 crypto exchanges have been offering perpetual and inverse perpetual contracts\footnote{If the base/numeraire currency of the future is switched, e.g. from \$ to BTC, then one uses the moniker `inverse'.} on Bitcoin and other tokens. 
Before we discuss the modifications introduced in the crypto-market, a short review of Shiller's original idea of extending the daily settlement mechanism beyond the change in future prices by including an extra term, which anchors the traded perpetuals to a potentially non-traded index. Conventional futures with the daily settlement price $F_t$ and a finite duration have a physical or cash settlement at the expiry date, whereas day-to-day price changes of the form $F_{t} - F_{t-1}$ link directly to matching cash flows between the counterparties.  
Since there is no expiry date for perpetuals a  method needs to be found to link the futures price to the index value. This is achieved by adding an extra term of the form $d_t - r_t F_{t-1}$. The dividend payments $d_t$, e.g. lumpy rental payments, are balanced with the riskless return $r_t$, e.g. a short-term risk-free rate associated with government debt, times the notional of the index. This additional daily exchange can be understood as representing the payout associated with owning the index minus the funding cost for such a position.

A similar methodology for linking flows between longs and shorts applies to perpetual crypto-currency futures.  The conventional daily settlement interval is reduced to 8 hours on BitMex and other periods on other exchanges. In addition, there is the challenge of defining a meaningful anchor to link the futures price to an index.  The first question is the choice of appropriate index. This is solved by choosing the Bitcoin spot rate either of the exchange where the perpetual is listed or some other related rate. Next, one has to associate a dividend $d_t$ with the index as well as define the respective interest rate $r_t$. Crypto-currencies do not pay dividends. 
BitMex overcame this difficulty by choosing a different kind of anchor, which in first approximation, when futures and spot price significantly deviate\footnote{This is the case, when the percentage difference between futures and spot reflected in $Premium\,\, Index(P )$ is significantly larger in absolute terms than $0.05\%$, i.e. when the `clamp' function in: $$Funding\,\,Rate (F) = Premium\,\, Index(P )+ clamp(Interest Rate (I) - Premium Index (P), 0.05\%, -0.05\%) $$ applies and $F \approx P$. The future price used is itself adjusted in a `unique' way to avoid price anomalies and avoid `unnecessary liquidation' according to the BitMex website. One should keep in mind that with maximal leverage of the kind offered, liquidation is unlikely to be an uncommon occurrence. 
The authors are not convinced, if in this context `unique' warrants an unqualified recommendation.} is proportional to the percentage difference between spot and a suitably adjusted futures price. Deribit launched perpetuals in 2018 with a simpler formula for the anchor\footnote{ For contract specifications see the Deribit website, \url{www.deribit.com}. }. As in the original perpetuals, the main payment for each period is due to the price change of the futures themselves. The added anchor payment per futures contract is defined as the product of funding rate and time fraction, where the time fraction is again divided into eight-hour slices, and the funding rate is given by  
$$Maximum (0.05\%, Premium ~ Rate) + Minimum (-0.05\%, Premium ~ Rate)$$ and the premium rate is given by 
$$Premium ~ Rate = ((Mark ~ Price - Deribit ~ Index) / Deribit ~ Index) \times 100\%$$where the Deribit index is the spot rate at the relevant time and the `Mark Price' corresponds to the relevant futures price adjusted to take into account bid-offer spreads and protect against potential market distortions. The other two markets, Binance and Bybit, have their own  formulas for aligning perpetuals with spot, i.e. with details to be found on the exchange websites. 

The benefits to speculators of trading in futures over spot is the additional leverage. The  initial margin for perpetuals on BitMex is only $1\%$  complemented by a maintenance margin of $0.35\%$ . This converts into high double digit leverage, which seems ridiculously high, since  double digit percentage daily moves are not uncommon.  This hyper-charges profits as well as losses and is particularly attractive for high-conviction momentum traders and as importantly the exchange itself. Consequently, when simple momentum trading signals based on moving averages are strong, the perpetual outperforms the spot since speculators are willing to pay to access leverage. The figure 6  shows the often lengthy periods of premiums and discounts between perpetual futures and spot. This  enabled hardy investors to accumulate return while being exposed to adverse moves in the underlying. 

\begin{figure}[h]
\includegraphics[width=\columnwidth]{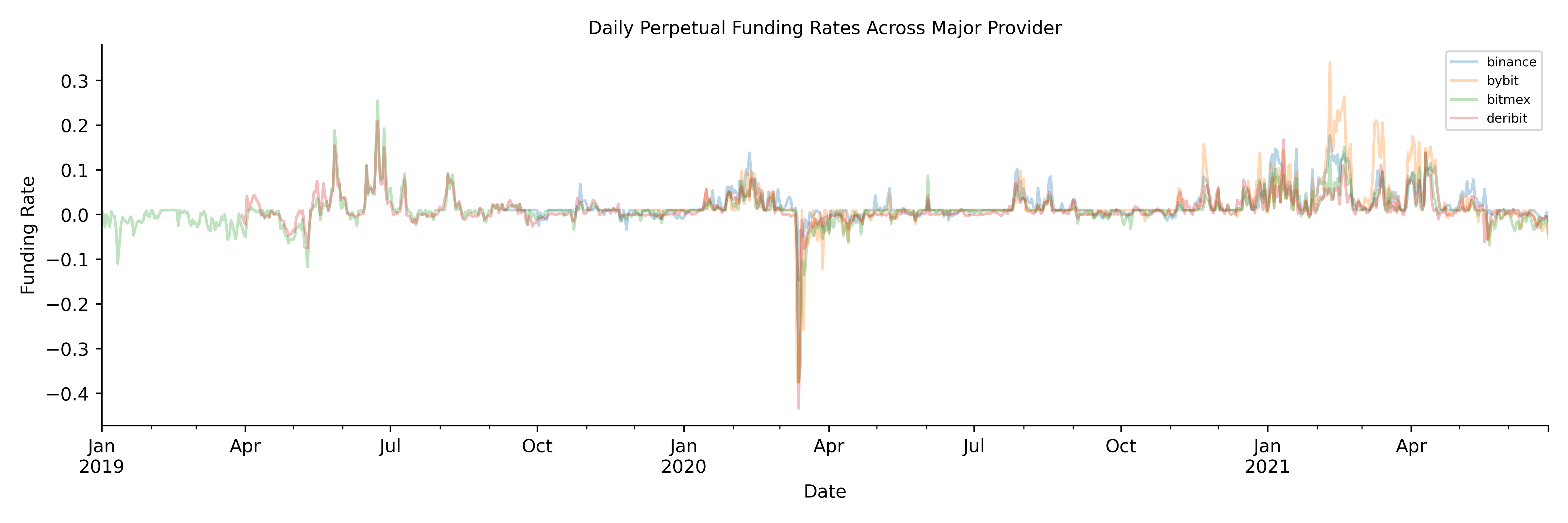}
\caption{  Funding Rates of Perpetuals in Percent on Multiple Exchanges. Data Source: cryptoquant.com}
\end{figure}

\subsection{The Basis of Finite Duration Futures} 
\noindent
A short section about interest rates implied from cryptocurrency futures of different maturities.
In figure 7 
the difference on BitMex, an innovator in the area of crypto-futures, between the perpetual and the shortest active future contract 
is plotted. 
During the run up of the Bitcoin price in 2020 and early 2021 the implied interest rate was consistently positive across many crypto-exchanges.  The futures basis on the  Chicago Mercantile Exchanges (CME) exhibited the same behaviour. Various explanations have been proffered. It is likely that restrictions on the availability of leverage played a role in the creation of a positive basis. Futures positions are more efficient than positions in spot, since many of the more popular exchanges (Binance, Deribit and Huobi for example) only require an initial margin of between 20- 50\% for futures, whilst spot positions  require full funding.
As a result a spread developed between spot and futures of different maturities. Short dated futures are cheaper to align with the spot price than long dated futures, which require a longer commitment of capital with its associated cost. This in itself leads to an upward sloping futures curve, but is maybe not an exhaustive explanation. Another possible reason is the difficulty for longer dated borrowing in cryptocurrencies due to the lack of general leverage and the limited connectivity between cryptofinance and the  wider capital market. 

\begin{figure}[h]
\includegraphics[width=\columnwidth]{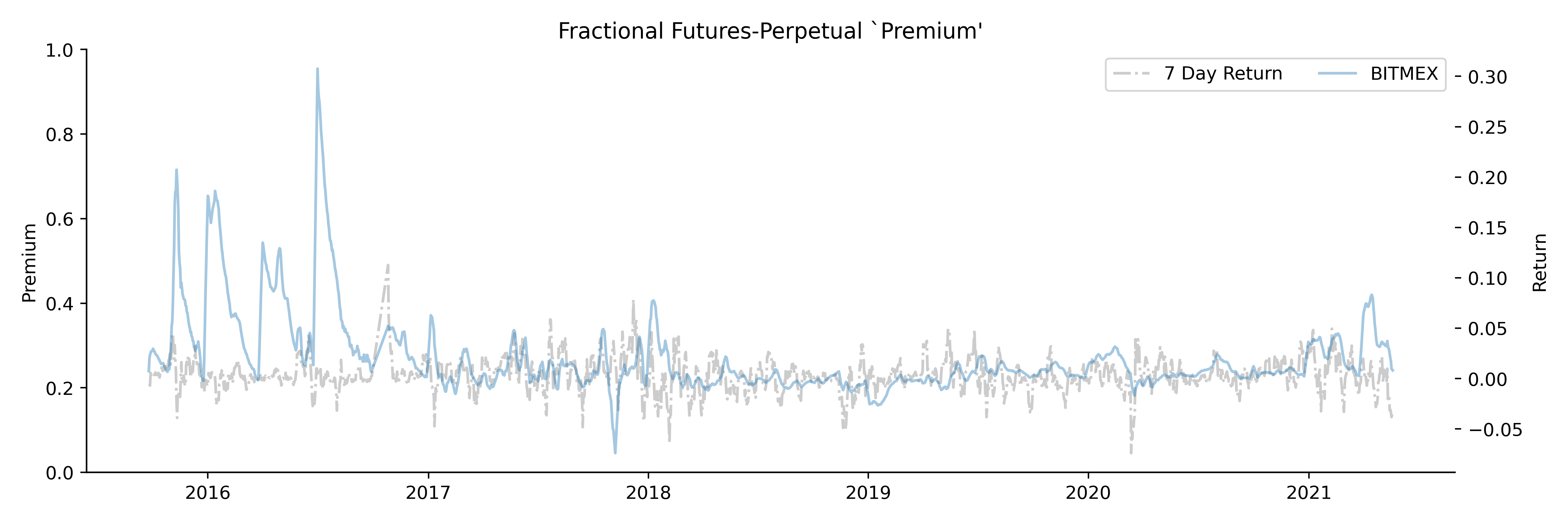}
\caption{Weekly rolling average of the   Premium between the perpetual and the closest futures contract to expiry: $\frac{F(t) - Perp(t)}{Perp(t)}$}
\end{figure}

\section{ Crypto-Options Market and  Yield }
\noindent
\begin{figure}[h]
\includegraphics[width=\columnwidth]{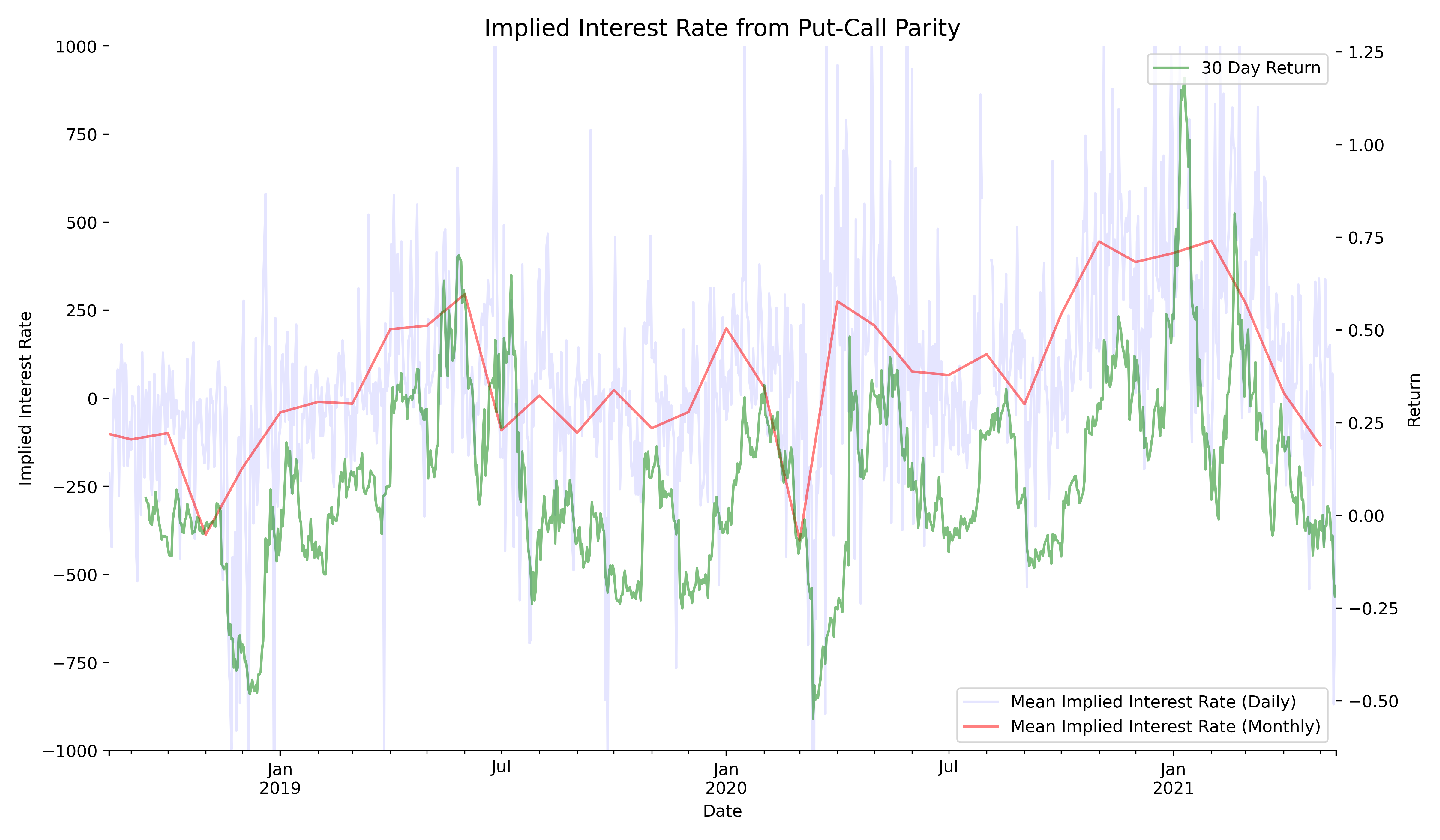}
\caption{  Put-Call parity implied interest rate, in annualised percent, from Options compared to the 30-day return of the underlying perpetual. Source: Deribit}
\end{figure}\noindent
This is another short section on CEX traded products. Implied interest is derived from the Derbit options market, the largest crypto-option market, using  put-call parity\footnote{The put-call parity is $B_t =   - \frac{ C_t - P_t  -  S_t}{K} $, where $C_t$ is the call price, $P_t$ the put price and $S_t$ the price of the perpetual contract, which acts as an underlying. The implied interest rate is $r   = - \frac{\ln{B_t}}{ T- t }$. It is calculated  expiry by expiry and strike by strike for a given day and then re-sampled, and the mean is taken across all instruments. $T- t$ is the time to expiry.}. All expiries and strikes are used, since the liquidity of individual options is  patchy, and both the weekly as well as calendar month average is plotted. The option implied interest rate, especially recently, plotted in figure 8 follows a similar dynamic as the interest rate associated with the futures in the last section. Implied yield, including convenience yield, has been discussed in Geman {\it et al.}\cite{geman2019,gemanprice2020} for Derbit options and CME futures.

In Alexander {\it et al.}\cite{alexander2021inverse} much is made out of the fact that not just futures but also options on the Deribit exchange are of the `inverse' variety. We cannot fully share this excitement. Inverse options, i.e. options were the numeraire currency is in Bitcoin but the notional amount is in USD, can be related by put-call symmetry to  conventional options in the Black-Scholes valuation.

\section{Conclusion:  Challenges of Decentralised Finance are Tied to Yield}
\noindent

Although the abstract stated that positions on the blockchain have no intrinsic yield, i.e. do not pay interest, this chapter described several interest-bearing crypto products. How can this apparent contradiction be resolved? Unencumbered by obligations and constraints, blockchain positions are instantly available, bypassing block size and updating constraints, and are accompanied by a convenience yield comparable to cash in fiat currencies\footnote{As a caveat, in mathematical finance incorporating cash is a challenge, and  similarly pure blockchain positions, if one ignores the somewhat artificial catch-all of convenience yield, pose a modelling difficulty.} as compensation for forgone interest payments.

Yields in traditional finance and cryptofinance can vary significantly\cite{disrupt_finance_2021_2}. This is partly because of market segmentation imposed by technological, regulatory, and institutional barriers and partly because yields are commensurate with risk. For instance, stable coins pegged to the USD offer significantly higher yields on DEXs and CEXs than deposits in traditional finance. This is because of varying risk profiles. While bank deposits are frequently insured, stable coins - particularly the dominant Tether - are viewed with suspicion for the possibility of breaking their pegs\footnote{In an  episode aired on  16\textsuperscript{th} June 2021 the stable coin Iron of Iron.Finance mostly but not completely backed by USDC and reliant for stability on minting the associated token Titan broke its peg. When the price of Titan fell after an earlier rapid appreciation, a gap opened up between the price on Sushiswap, a decentralised exchange (DEX), and the lagging TWAP
 (`time weighted average price'),
 which functioned as the oracle for the stability mechanism. As a consequence, the incentive system designed to support Iron when the coin was below \$1 by allowing arbitrageurs to buy Iron  in the market and then redeem it at the full value as a mix of USDC and Titan failed, since the market discount in Iron was insufficient to compensate for the difference between the TWAP and the rapidly vanishing market value of Titan.  Without the buying support of arbitrageurs Iron failed to recover and   Titan fell further, as faith was lost in the mechanism. This lead to a vicious spiral with  ever increasing price reductions in Titan driving the value from over $\$ 60$ to close to zero.  The TLV tumbled  within a day  by $99+\%$ from a peak of over $\$ 2$ billion. For a fuller analysis one has to also consider minting, i.e. the creation of Iron, and more importantly the motivation of Titan sellers. A good analysis, including market data for the critical day, can be found on the blog of I. Kuznetsov\cite{kuznetsov_2021}.}.

The introduction of staking, which enables the pledging of on-chain positions using token-specific mechanisms, has ideally created something strikingly similar to the short-term rates provided by central banks for fiat currencies. This affects the dynamics of decentralised finance, and the ramifications will only become clear over time.

The chapter's centrepiece  
are various yield-generating products and strategies that are currently popular for cryptocurrencies. The majority of these are less than five years old, and several have only recently gained prominence, with their TLVs reaching billions within months. Due to their complexity and the market's byzantine structure, each of them provides ample opportunity for additional research. We presented novel data from DEXs, CEXs, and the Ethereum 2.0 staking protocol. The chapter illustrates  
that cryptofinance is still grappling with leverage and yield enhancement.

Iterative development is used to create DeFi applications. Evolutionary improvement occurs through successive generations. Flawed approaches are forked away and vanish in an ideal world before any `damage' is done. The term `damage' is relative, and billions of dollars worth of tokens have already been lost due to various failures\footnote{A recent episode involved the Uranium (a fission bomb) project, which reached a critical mass of \$50 Million and blew up,
because a security flaw in the smart code went unnoticed.}. There could be a `winner-takes-all' outcome, with the vast majority of attempts failing quickly and spectacularly, or there could be cycling through various phases with no stable tokens or DEXs\footnote{Tokens and DEXs could also be confronted with an unending series of forks in their technology development path. Any potentially hard-to-reverse sub-optimal choice throws them out of the race since a multitude of competitors can spring up, copy all their correct historical decisions and overtake incumbents. Reputation or legitimacy mentioned as a way to protect a leadership position might not suffice, since as traders know, `one is only as good as the last trade'.}. The function of tokens as `means of exchange' may preclude the existence of an equilibrium solution. Rather than `failure being an impossibility', failure should be considered the default outcome.

Cryptofinance is embedded in economic history. Finance is littered with ostensibly limitless wealth schemes. The Mississippi and South Sea bubbles, which peaked in the early 1720s and were associated with France and Britain, are particularly well-known. In France, John Law, an enterprising Scotsman, on the rise and possibly on the take, convinced the court to grant the Mississippi company a broad monopoly in exchange for the country's debt being assigned to a company-linked bank. This captured the public imagination, and the company's market value briefly exceeded France's GDP\footnote{The joint crypto market capitalisation is only about 2\% of world GDP, and by this measure could still have some room to grow.} before collapsing. This inauspicious
affair fostered a (un)healthy scepticism of financial innovation. Whereas the South Sea bubble did not have the same effect on financial markets in Britain. Finance in its various forms, from double-entry bookkeeping to access to the London capital market, underpinned the East India Company's success, which for a time-controlled a large portion of the Indian subcontinent, accounted for half of the world trade, was involved in large-scale famines such as the one in 1770, and maintained a standing army of over 250,000. The moral of the storey: Getting finance right is critical. For the same reason, a sound framework and finely tuned parameters\footnote{This is not easy, and even the designers of the board game Monopoly needed more than a decade to come up with amenable
rules.} will be critical for DeFi to enable leverage and sustainable yields.

Cryptofinance's Galapagos syndrome with rich innovation isolated from traditional finance may be coming to an end, as high profile institutions invest in the more vanilla end of the crypto spectrum, mostly Bitcoin \footnote{\url{https://bitcointreasuries.net/}}. Corporations like MicroStrategy and Tesla and a host of banks are notable examples. Terms such as Hybrid Finance (HyFi) are gaining prominence as traditional players and regulators circle the new asset class and stake claims in cyberspace as they see fit.

Institutional involvement is not necessarily at odds with the original  premise of crypto. Namely that legacy finance provides insufficient yield and is inherently unstable. This is already enshrined in the bitcoin's genesis block \footnote{\url{https://en.bitcoin.it/wiki/Genesis_block}} as, `The Times 03/Jan/2009 Chancellor on brink of second bailout for banks'. 
How long can  institutions resist the siren call of alternative yields in cryptofinance?

\noindent We gratefully acknowledge conversations with DC Brody and Sreejith Das.

\bibliography{bibfile.bib}

\end{document}